\documentclass{nature}
\usepackage{amsmath}
\usepackage{amsfonts}
\usepackage{graphicx}
\usepackage[dvipsnames]{xcolor}
\usepackage{color}
\usepackage{multirow}
\usepackage{hyperref}
\usepackage{adjustbox}

\usepackage{caption}
\usepackage{float}
\makeatletter
\let\saved@includegraphics\includegraphics
\AtBeginDocument{\let\includegraphics\saved@includegraphics}

\makeatother

\title{Quantifying the relationship between specialisation and reputation in an online platform}

\begin{document}
\maketitle


\author{Giacomo Livan,${}^{1,2,*}$ Giuseppe Pappalardo,${}^{3}$ Rosario N. Mantegna${}^{4,5}$}

\begin{affiliations}
 \item Department of Computer Science, University College London, London WC1E 6EA, United Kingdom
 \item Systemic Risk Centre, London School of Economics and Political Sciences, London WC2A 2AE, United Kingdom
 \item Dipartimento di Fisica `Ettore Majorana', Universit\`a di Catania, Via S. Sofia, 64, 95123 Catania, Italy
 \item Dipartimento di Fisica e Chimica Emilio Segr\`e, Universit\`a di Palermo, Viale delle Scienze, Ed. 18, 90128 Palermo, Italy
 \item Complexity Science Hub Vienna, Josefst\"adter Strasse 39, 1080, Vienna, Austria
\end{affiliations}

\begin{abstract}
Online platforms experience a tension between decentralisation and incentives to steer user behaviour, which are usually implemented through digital reputation systems. We provide a statistical characterisation of the user behaviour emerging from the interplay of such competing forces in Stack Overflow, a long-standing knowledge sharing platform. Over the 11 years covered by our analysis, we find that the platform's user base consistently self-organise into specialists and generalists, i.e., users who focus their activity on narrow and broad sets of topics, respectively.  We relate the emergence of these behaviours to the platform's reputation system with a series of data-driven models, and find specialisation to be statistically associated with a higher ability to post the best answers to a question. Our findings are in stark contrast with observations made in top-down environments — such as firms and corporations — where generalist skills are consistently found to be more successful.

\end{abstract}

The evolution of the digital economy has transformed several top-down online environments into bottom-up, decentralised platforms. For instance, information and news are now largely consumed via social media~\cite{gottfried2016news}, and well established business-to-consumer sectors -- such as the hotel industry~\cite{zervas2017rise} -- have been disrupted by sharing economy platforms, which empower users to become small entrepreneurs by sharing spare resources.

Because of their decentralised nature, over the years several online platforms have introduced a variety of incentive systems to foster trust between their users. In some cases (e.g., Twitter), these come as simple identity verification protocols. In other cases (e.g., sharing economy platforms such as Uber and Airbnb), trust is fostered with a reputation score that users develop through digital peer-review mechanisms (e.g., star ratings)~\cite{tadelis2015economics,tadelis2016reputation}.

In recent years, a number of studies have analysed the relationship between user behaviour and reputation. Experimental approaches have measured user response to different elements appearing on profiles in order to identify which ones are most conducive to trust~\cite{zloteanu2018digital}. Other studies have instead looked at strategic behaviour as a driver of user reputation, focusing, e.g., on the the cooperative and retaliatory mechanisms underlying the exchange of ratings~\cite{livan2017excess,zervas2021first} and on the incentives to commit review fraud~\cite{luca2016fake}. An understudied aspect in this stream of research relates to other types of strategic user behaviours, namely those related to specialisation and/or generalism.

In ecology, the term specialist (generalist) refers to species that prosper in a limited (wide) range of environmental conditions. Specialisation emerges as a natural response to competitive pressure, with the aim of securing an edge in specific circumstances. Conversely, generalism emerges as resilience against varying conditions. Such concepts have found plenty of applications in non-natural domains, and have been particularly helpful to conceptualise different strategic behaviours in large socio-economic systems. 

The management literature has consistently found that individuals with broader sets of skills (i.e., generalists) enjoy greater success in top-down organisations. Generalist CEOs receive higher pay than their specialist counterparts, with the highest pay increases occurring when firms switch from a specialist to a generalist CEO~\cite{custodio2013generalists}. Similar results are found in~\cite{brockman2016determinants}, which the authors interpreted as a reflection of a higher demand for generalist skills required to manage increasingly complex firms, and generalist CEOs are more likely to engage in acquisitions outside a firm's main industry~\cite{chen2021generalist}. Similarly, empirically tested theories of leadership support the idea that leaders in industry tend to be generalists rather than specialists~\cite{lazear2012leadership}.

Traces of such behaviours have been also observed in the bottom-up context of online platforms, with a wide range of strategies -- ranging from extreme specialisation to extreme generalism -- being found, e.g., on Reddit and GitHub~\cite{waller2019generalists,onoue2013study}. Notably, such strategies are associated to different user archetypes, with specialists being more likely to stick to the online communities they contribute to and generalists being more likely to remain active on platforms as a whole. In the context of online gaming, generalists have been found to be more resilient to change (e.g., after the release of game patches) although specialists ultimately tend to outperform other players on average~\cite{jiang2021wide}.

In this paper, we aim to quantify the relationship between specialisation/generalism and reputation in online platforms. To the best of our knowledge, this is an understudied relationship, which has only been looked at in contexts where reputation is developed through interactions that are external to platforms (e.g., the online ratings received by medical professionals on physician-rating websites~\cite{pike2019online}). Our focus here, instead, is to look at such a link in contexts where user reputation is developed \emph{endogenously} through interactions and peer-review taking place on the platform itself. We do so by analysing data from Stack Overflow (SO), the flagship knowledge-sharing platform of the Stack Exchange network, which features questions and answers on a wide variety of topics in the area of computer programming (see Methods section). SO implements an elaborate reputation system, which is well known for its effectiveness in incentivising users to produce high quality posts~\cite{movshovitz2013analysis}. 

\section*{Results}

\subsection{Platform growth}

\begin{figure*}[h!]
\centering
\includegraphics[scale=0.6]{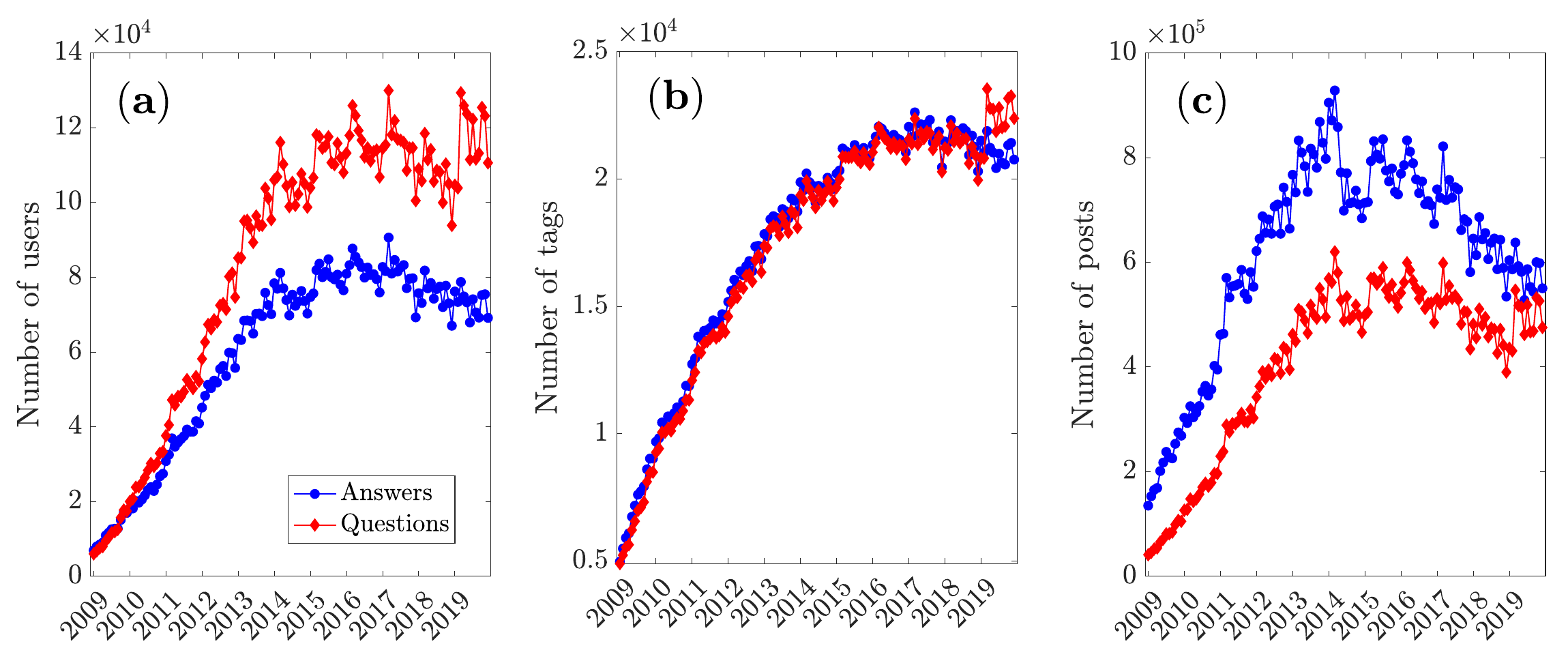}
\caption{\footnotesize{Growth of Stack Overflow from 2009 to 2019. $(\mathbf{a})$ Monthly number of active users. $(\mathbf{b})$ Monthly number of tags featured in posts. $(\mathbf{c})$ Monthly number of posts. In all panels the blue (red) symbols refer to answers (questions) on Stack Overflow.}}
\label{fig:growth}
\end{figure*}

We begin our analysis by looking at the evolution of the Stack Overflow platform over time from an aggregate perspective. For each year in our dataset (2009-2019), we look at the monthly number of active users (i.e., users who posted at least once), the monthly number of tags (i.e., tags that appear in at least one post), and the monthly number of posts. These quantities are reported in Fig.~\ref{fig:growth}, broken down by post type. The number of users posting questions rapidly overtakes the number of those posting answers (left panel), and both numbers settle around several thousands around 2013-2014. The number of tags featured in answers and questions roughly equal each other throughout the platform's lifetime (central panel), whereas the number of answers posted remains systematically higher than the number of questions (right panel), with both numbers settling on the order of tens of thousands of posts per month. 

We then proceed to characterise the platform's growth by categorising its user base with respect to post types. We discard casual users by restricting our analysis to those who contribute with at least 10 posts (answers and questions combined) in a given year. We characterise a user's activity based on the relative proportion of questions and answers. We indicate as $A_i(y)$ ($Q_i(y)$) the number of answers (questions) posted by user $i$ during year $y$, and we characterise the user's profile with respect to post types in that year with the following score:
\begin{equation} \label{eq:activity_score}
D_i(y) = \frac{A_i(y)-Q_i(y)}{A_i(y)+Q_i(y)} \ .
\end{equation}

\begin{figure*}[h!]
\centering
\includegraphics[scale=0.5]{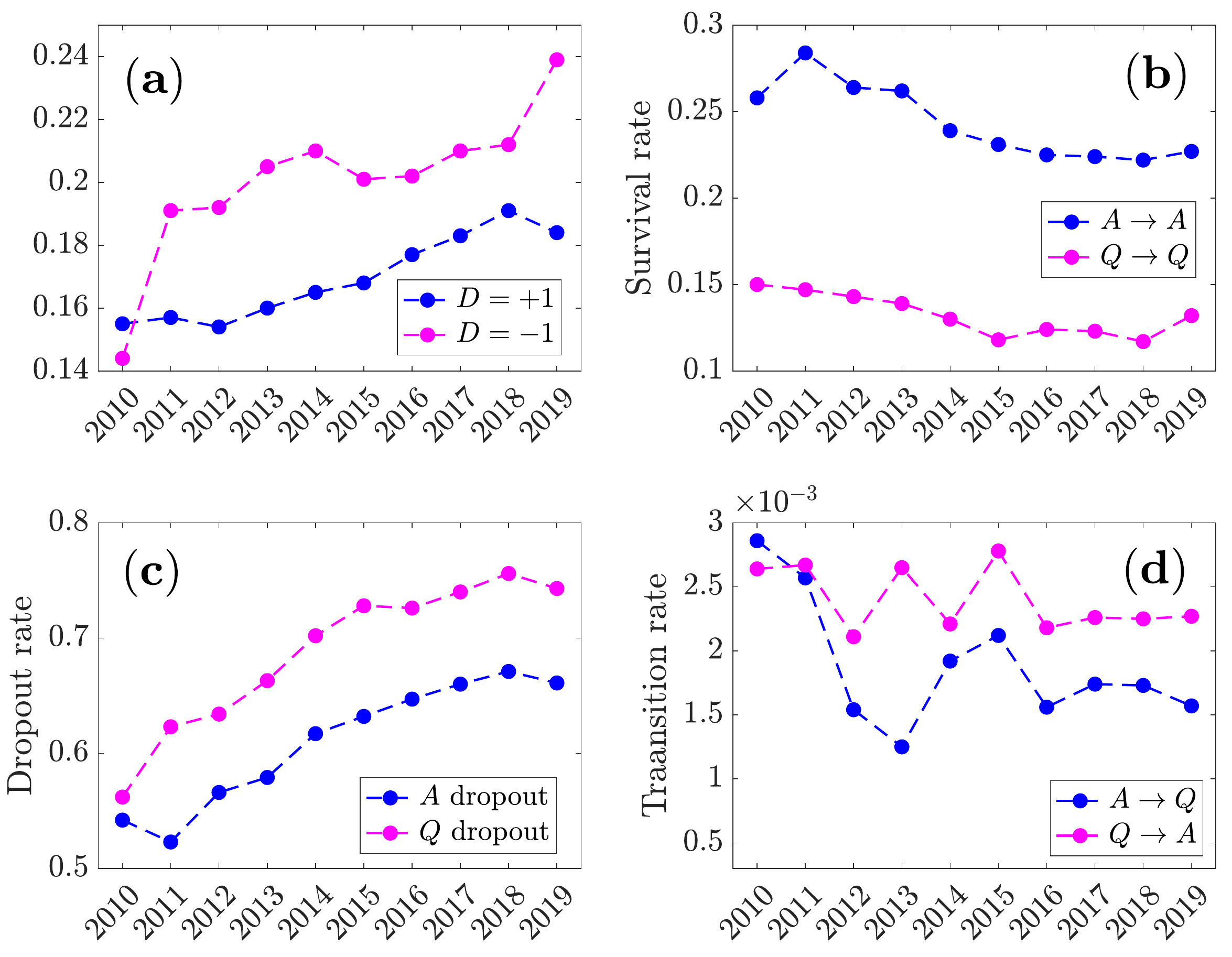}
\caption{\footnotesize{Characterisation of Stack Overflow's user base. $(\mathbf{a})$ Annual percentage of $A$- and $Q$-users ($D=1$ and $D=-1$, respectively, see Eq.~\eqref{eq:activity_score}). $(\mathbf{b})$ Annual survival probabilities for $A$- (blue) and $Q$-users (magenta), defined as the empirically estimated probabilities for users belonging to either group to belong to the same group in the following year. $(\mathbf{c})$ Annual dropout rates for $A$- (blue) and $Q$-users (magenta), defined as the empirically estimated probabilities for users belonging to either group to either leave the platform or fall below the minimum threshold of 10 posts per year to be considered in our analysis. $(\mathbf{d})$ Annual transition rates from $A$- to $Q$-users (blue) and vice versa (magenta), defined as the empirically estimated probabilities for users belonging to one group to transition to the other one the following year.}}
\label{fig:survival_dropout}
\end{figure*}

Figure~\ref{fig:survival_dropout} (top left) shows the annual proportions of users who only post answers ($D_i = +1$) or questions ($D_i = -1$). Let us label such two groups as $A$-users and $Q$-users, respectively. Overall, the proportions of users belonging to both groups grow over time. However, the fraction of $A$-users remains relatively stable between $15\%$ and $20\%$ (even showing some decline in 2019), whereas the proportion of $Q$-users increases from less than $10\%$ to almost $25\%$. After an initial phase where $A$-users are more numerous, $Q$-users become the relative majority in 2011, signalling the platform's transition from a `supply-driven' to a `demand-driven' knowledge marketplace.

The above transition is not driven by the addition of newcomers to a stable core of users, but rather by turnover. The top right and bottom left panels in Figure~\ref{fig:survival_dropout} show -- respectively -- the year-to-year survival and dropout rates for $A$- and $Q$-users. With the former, we indicate the empirically estimated probability that a user belonging to either group in a given year will again belong to the same group the following year, while with the latter we indicate the probability that a user either leaves the platform or falls below the minimum activity threshold to be included in our analysis (10 posts).  Only a minority of $A$- and $Q$-users remain in such groups in consecutive years, and the dropout rates for both groups display a sharp increase over time. We can therefore conclude that the sub-populations of $A$- and $Q$-users grow over time through the replacement of users who drop out with larger numbers of new users. 

Let us also mention that there is very little spillover between the two groups throughout the years, as testified by the fact that the transition rates between them (i.e., the empirically estimated probability that a $Q$-user will become an $A$-user the following year and vice versa) are both below $0.3\%$, as shown in the bottom right panel in Figure~\ref{fig:survival_dropout} .

\subsection{Specialist and generalist users}

We then proceed to characterize user behaviour in terms of topics. We do so by forming monthly bipartite user-tag networks restricted to `pure' $A$- and $Q$- users (i.e., users whose activity score in Eq.~\eqref{eq:activity_score} is $D = 1$ and $D = -1$, respectively, in the year of interest). Namely, if a $Q$-user $i$ has posted $w_{i\tau}^Q$ questions featuring the tag $\tau$, we place a link from $i$ to $\tau$ with weight $w_{i\tau}^Q$. We construct a similar network for $A$-users, considering as weights the number of answers posted in response to questions featuring a certain tag. Following well established approaches to detect the coexistence of specialisation and generalism in ecosystems, we measure \emph{nestedness} in such networks (see Fig.~\ref{fig:nestedness}), and compare its values against those obtained under a null network model in order to establish its significance, following a procedure based on spectral radii~\cite{staniczenko2013ghost} (see Methods section). We find nestedness to be statistically significant throughout the platform's history (see  Fig.~\ref{fig:nestedness}), which in turn suggests that the platform indeed self-organises into specialist and generalist users, both in its supply and demand sides.

\begin{figure*}[h!]
\centering
\includegraphics[scale=0.5]{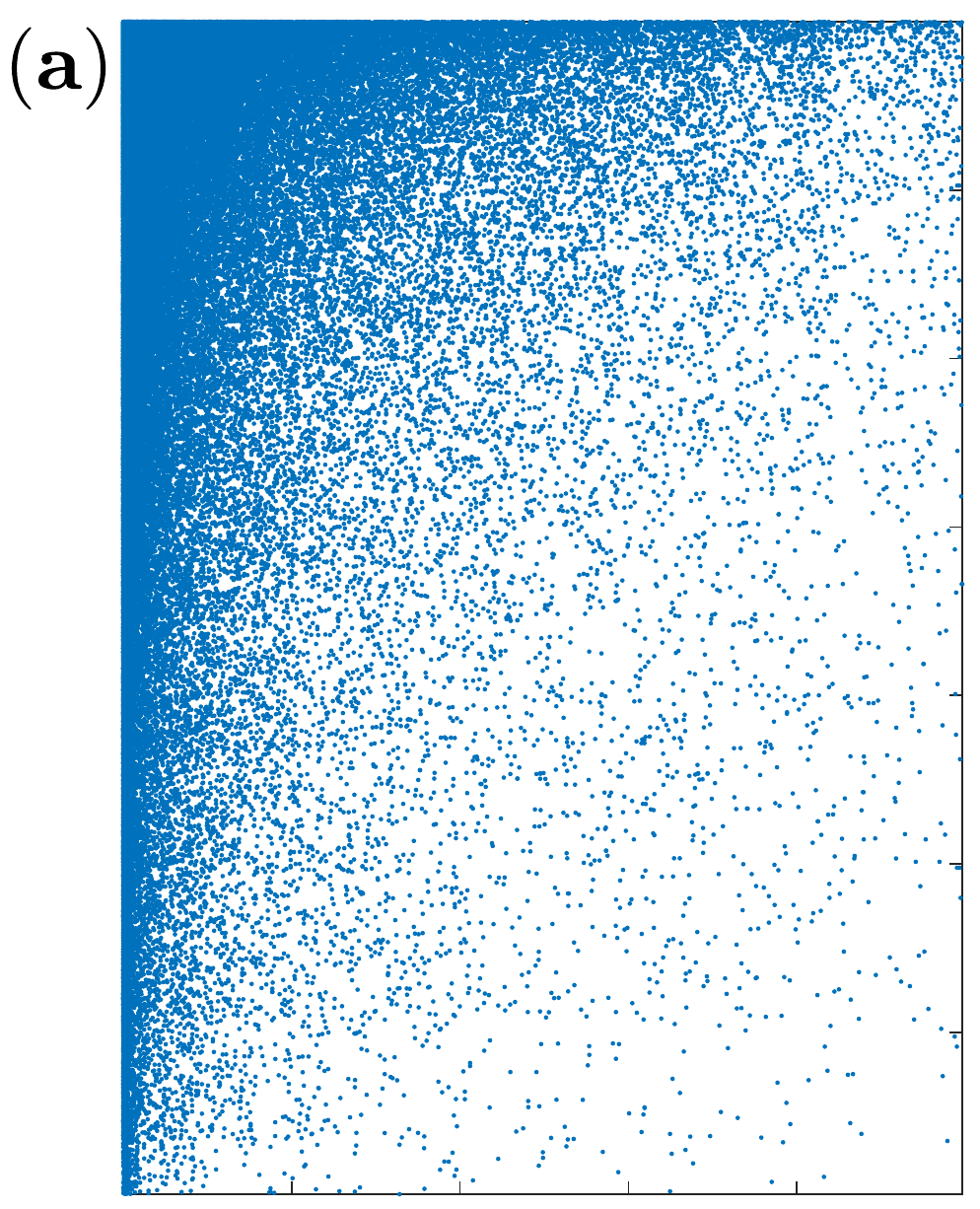}
\hspace{0.05\textwidth}
\includegraphics[scale=0.5]{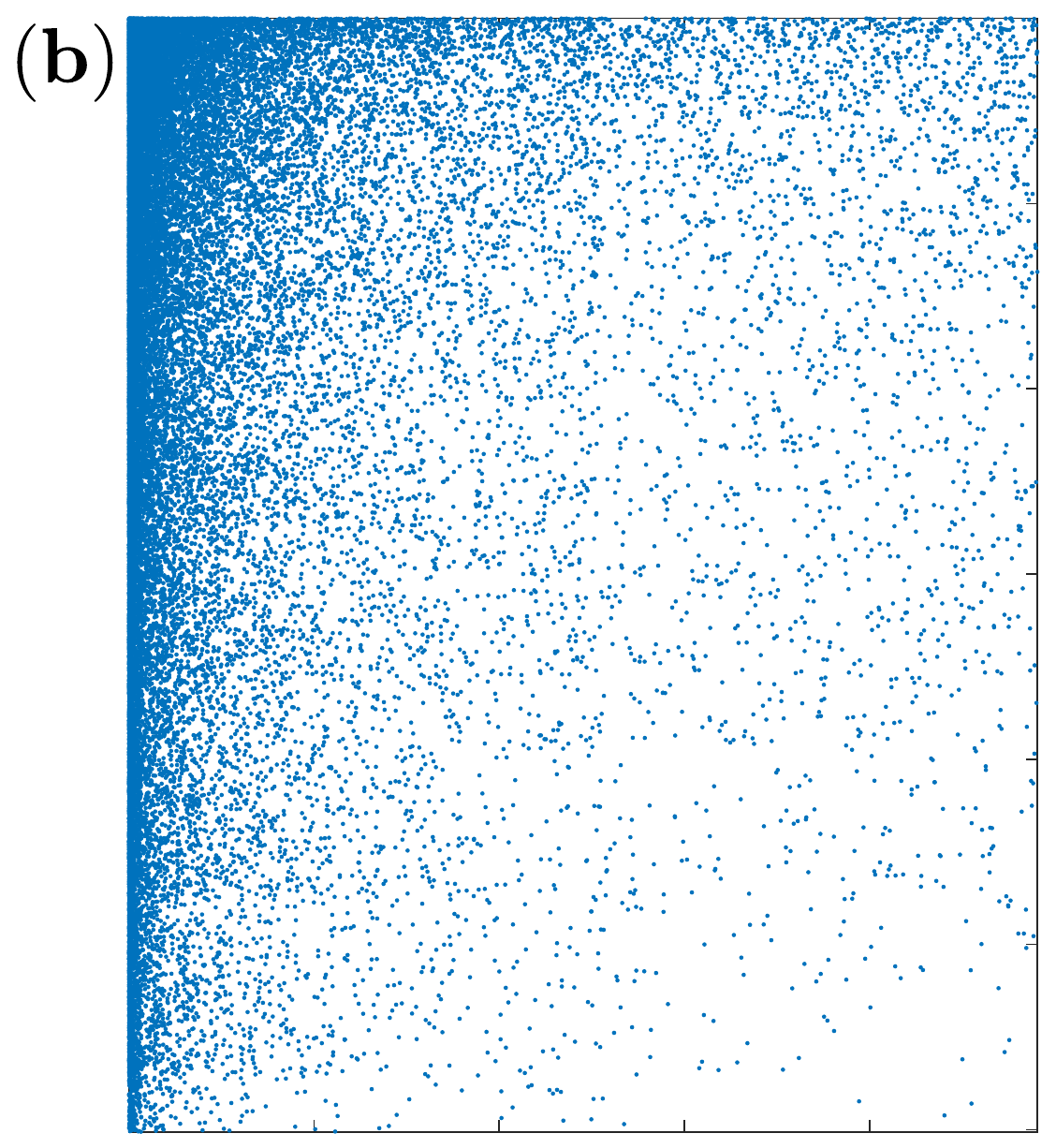}
\caption{\footnotesize{Evidence of nestedness in Stack Overflow's user-tag bipartite networks. $(\mathbf{a})$ User-tag bipartite network for answers posted in January 2009 (size: $6960 \times 4982$). $(\mathbf{b})$ User-tag bipartite network for questions posted in January 2009 (size: $6011 \times 4906$). In both panels blue dots represent non-zero entries, and the matrix rows and columns have been sorted from top to bottom for users and from left to right for tags.}}
\label{fig:nestedness}
\end{figure*}

Based on this observation, we then quantify the level of specialisation attained by users in their activity when posting answers/questions with the Herfindahl index, a measure of concentration which (in the case of questions) reads
\begin{equation} \label{eq:herfindahl}
H_i^Q = \sum_\tau \left ( \frac{w_{i\tau}^Q}{s_i^Q} \right )^2 \ ,
\end{equation}
where $w_{i\tau}^Q$ (as defined above) is the number of questions posted by the user on tag $\tau$, whereas $s_i ^Q= \sum_\tau w_{i\tau}^Q$ is the total number of questions posted by the user. With the above definition, the Herfindahl index will approach one for users who are only active on a limited set of tags (with the limiting case $H_i^Q = 1$ for users active on just one tag), and will instead approach zero for users whose activity is uniformly spread  over a large number of tags. We define an equivalent index $H_i^A$ in the case of answers and characterise users whose activity features both types of posts with both Herfindahl indices. 

Fig.~\ref{fig:Herfindahl_pdf} shows the annual distributions of the Herfindahl scores for both answers and questions. Both distributions are remarkably stable throughout the years, signalling that -- despite the increase in the number of tags (see the middle panel in Figure~\ref{fig:growth}) -- the users' collective behaviour in terms of specialization remains largely unchanged.

\subsection{Reputation}

We proceed next to investigate the users' reputation in the platform. For each user with at least 10 posts in a year,  we build a profile based on the following features describing their activity: the number of posts ($n$), the number of tags associated to their posts ($t$), their Herfindahl indices ($H^A$ and $H^Q$, see Eq.~\eqref{eq:herfindahl}), and their activity score ($D$, see Eq.~\eqref{eq:activity_score}). We use these features to build a number of linear models to characterise user reputation on the platform. 

\begin{figure*}[h!]
	\centering
	\includegraphics[scale=0.5]{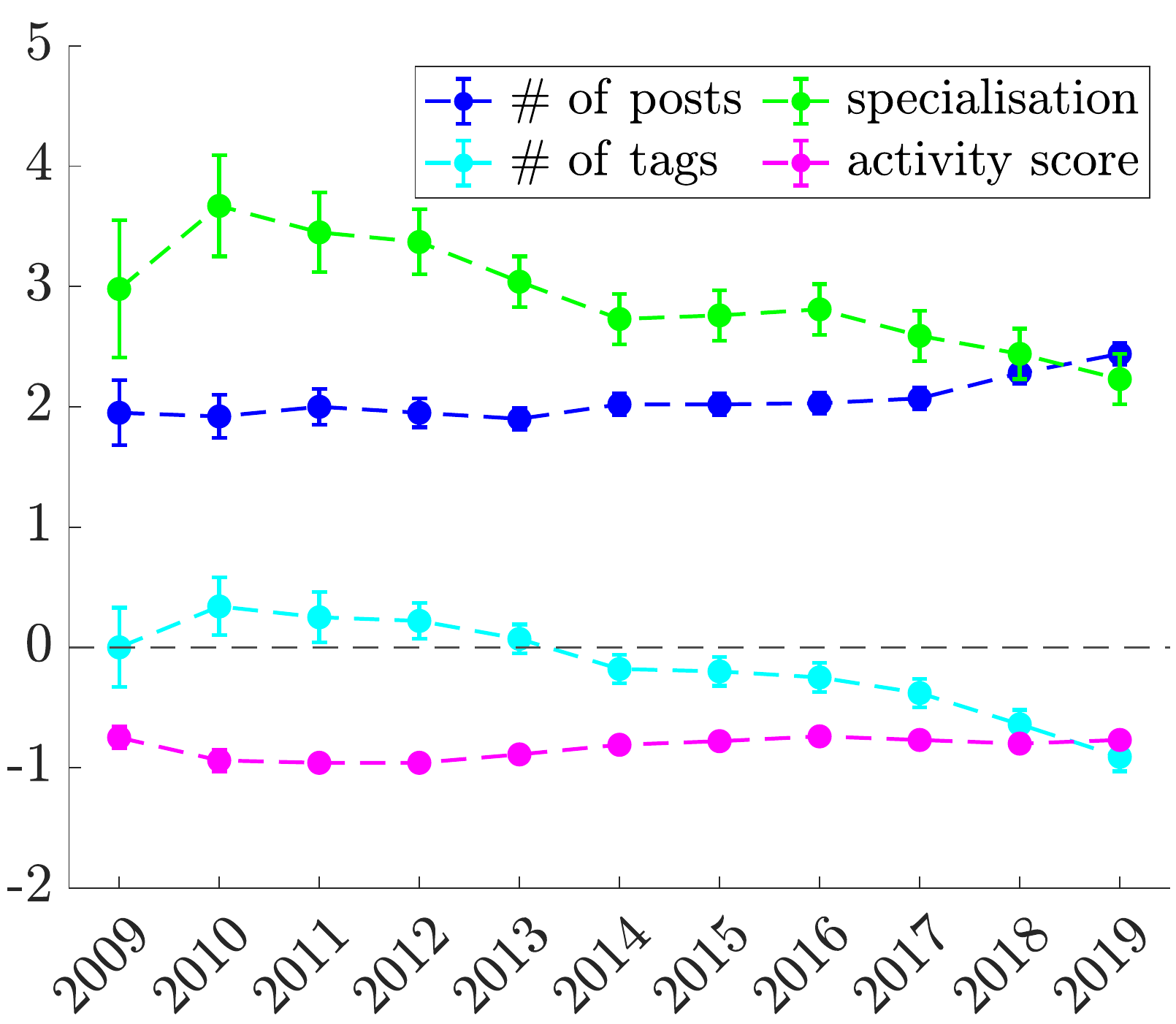}
	\caption{\footnotesize{Logistic regression results for the probability that a user has at least one accepted answer in a given year. Dots represent the values of the regression coefficients estimated for the four covariates included in the model, shown in the legend. Error bars show the standard errors on the coefficients times three.}}
	\label{fig:regression_coeff}
\end{figure*}

We begin by looking at the main sources of user reputation, i.e., the ability to post \emph{accepted} answers. These correspond to answers selected as the best one in response to a given question by the author of the very same question. Notably, posting an accepted answer is worth 15 reputation points (whereas an up-vote, for instance, is worth 10), and it is the result of a combination of skills (i.e., both competence and rapidity). In order to identify the factors that are conducive to a user's ability to post answers that may get accepted, we consider a logistic regression model for the log-odds $\log (\pi_a / (1-\pi_a ))$, where $\pi_a$ denotes the probability of a user having at least one accepted answer in a given year (see Methods section). We choose to do so -- instead of modelling the acceptance rate of a user's answers -- because we find the user population to be approximately split between those who have at least one accepted answer and those who have none. The full results of the calibration of the above model are shown in Table~\ref{tab:regression_acceptance}, with the corresponding ROC curves shown in Fig.~\ref{fig:ROC}. Throughout the years, the model delivers excellent accuracy (with an AUC ranging between $76\%$ and $84\%$). The regression coefficients obtained for each covariate in each year of our analysis are illustrated in Fig.~\ref{fig:regression_coeff}. For the first ten out of eleven years, specialisation ($H^A$) is found to be the leading contributor to a user's ability to post high-quality answers, reaching its maximum relative importance in the early years of the platform, with some mild decline in more recent years. 
User activity ($n$) is the second main contributor, with an increasing trend suggesting that it may be overtaking specialisation (albeit the two coefficients are statistically indistinguishable both in 2018 and 2019). Notably, the number of tags $t$ on which a user posts answers is the only covariate whose impact changes over time: before 2013-2014 it contributes positively to a user's ability to post accepted answers, while it contributes negatively to it after then. This somewhat further strengthens the importance of specialisation, as it suggests that the more successful users are those who specialise on narrower sets of tags. The activity score $D$ remains instead negatively correlated with the ability to post accepted answers throughout the entire time window of our analysis.

We then build a multinomial logistic regression model to classify users (in each year) into three mutually exclusive categories: users whose posts have received zero votes, users whose posts have received only up-votes, and users whose posts have received both up- and down-votes. We neglect the case of users whose posts only receive down-votes, since in all years considered in our analysis they are less than $0.1\%$. We  calibrate multinomial logistic regression models for the log-odds associated with the probability of belonging to the three above categories, using the aforementioned features as covariates. The full results are reported in Tables~\ref{tab:multinomial_A} and~\ref{tab:multinomial_Q},  and show that no specific feature is systematically associated with a higher probability of attracting votes.

We then proceed to restrict our analysis to those users whose posts received at least one vote in a given year. To this end, we calibrate four stepwise regression models using as dependent variables the (logarithm of the) average number of up- or down-votes per post received by a user. Starting from a constant model, we use both forward and backward selection to select the best model (in terms of sum of squared residuals) based on the aforementioned covariates. With only few exceptions, the stepwise selection procedure results in a very simple model where the users' activity -- as quantified by their number of posts $n$ -- is the only statistically significant covariate. However, it is noteworthy that activity has a similar impact across board, both in terms of sign and magnitude. Namely, we find activity to have a negative impact on the number of votes received per post, both in the case of up- and down-votes and regardless of the type of post. Remarkably, in the case of questions such minimalistic models explain 60\% or more of the variance. The full results of the calibration are reported in Tables~\ref{tab:stepwise_A_up}, \ref{tab:stepwise_A_down}, \ref{tab:stepwise_Q_up}, and \ref{tab:stepwise_Q_down}.

\section*{Discussion}

In this paper we presented a number of analyses aimed at understanding the relationship between specialisation and reputation in the domain of online decentralised platforms. Thanks to the lack of monetary incentives and its already long history, Stack Overflow represents an ideal environment to observe the development of such a relationship `in the wild' over an extended period of time.

The 11 years of history covered in our study reveal how the Stack Overflow platform's user base grew into a structured community, with different individuals taking on different roles. First, we documented how most of the platform's user base quickly evolved into well defined supply and demand sides, represented by two large sub-communities of users characterized by their willingness to answer or pose questions, respectively. Second, we provided ample evidence on the emergence of specialisation at the level of topic selection in the users' posts.   

Should the above findings be attributed to self-organisation or should they instead be interpreted as a direct response to the platform's design and incentives? Plausibly, the very nature of Stack Overflow -- a knowledge-sharing platform structured around questions and answers -- is responsible for the emergence of sub-communities dedicated to posting answers and questions. Like other two-sided platforms, Stack Overflow naturally attracts users with markedly different needs (e.g., similarly to hosts and guests in accommodation platforms). 

Specialisation with respect to topic selection is a more complicated phenomenon to unpack. We do not find it to be correlated (in a statistical significant manner) with the likelihood of attracting up- or down-votes to generic posts, suggesting that the quality of a user's posts may be largely idiosyncratic. Conversely, we do find a statistically significant correlation between a user's specialisation and the likelihood of their answers being accepted as the best one in response to a question. This is a notable asymmetry, as an equivalent selection mechanism is lacking in the case of questions, and no other user-generated feedback awards more reputation points on Stack Overflow than an accepted answer. We interpret these findings as a clear consequence of the incentives set in place by the platform's reputation system. 

It is interesting to relate our results to findings about the users' decision-making when choosing which answers to accept. Such decision-making has been found to be largely driven by heuristics, with selections being determined by factors such as the order in which answers appear or the amount of screen space they occupy~\cite{burghardt2017myopia}. It is therefore tempting to speculate that the selection process that takes place on posted answers may contribute to optimise user behaviour with respect to such heuristics. 

Our findings illustrated in Fig.~\ref{fig:regression_coeff} shed light on the above point by identifying the salient traits of successful users. These are -- on average -- highly active and specialised users, whose specialisation progressively focuses on a narrower set of topics (as testified by the change in sign of the coefficient associated to the number of tags). Notably, these are not users who specialise in posting answers only, as their activity score $D$ (see Eq.~\eqref{eq:activity_score}) is negatively correlated with the likelihood of having answers accepted, suggesting that developing some expertise on both sides of a two-sided platform may unlock positive reputational spillovers.

Overall, our findings are in rather stark contrast with observations made in top-down environments (such as firms and corporations), where generalists are usually found to enjoy greater success than specialists. However, we ought to acknowledge that the extent to which our findings may generalise to other decentralised online environments can only be the subject of speculation at this stage. Stack Overflow's reputation system and the sustained success it has brought to the platform -- with relatively minimal policy changes throughout the years -- are rather unique. Other successful knowledge-sharing platforms have taken radically different approaches to foster trust within their user base. For instance, Wikipedia holds elections to promote reliable users to administrators. Similarly, comparisons with different reputation/feedback systems (e.g, textual reviews) are not straightforward. We therefore believe our work represents a first step towards the data-driven modelling of the relationship between specialisation and online reputation, and a blueprint that following studies may adapt to different environments and data sources. 
 
\section*{Methods}
\subsection{Data}

We analyze data from the Stack Overflow platform, the flagship site of the Stack Exchange Network, which features questions and answers on a wide variety of topics in the area of computer programming. The portion of the data used in our study spans 11 years going from January 2009 (shortly after the platform was started in 2008) to December 2019.  Posts represent the main unit of activity in the platform. Posts are divided into three main categories: questions, answers, and accepted answers. An accepted answer is a post that has been identified as the best one in response to a question by the author of the same question. Users can classify the questions they post with up to five tags (e.g., C++, Python, etc), which help other users identify the posts they might be able to reply to. Each individual post (i.e., both questions and answers) can generate a sub-thread in the form of comments.  Any post or comment can be either up-voted or down-voted by other users. Users develop a reputation score based on their activity. The main source of points are accepted answers ($+15$ points) and up-votes ($+5$ for questions, $+10$ for answers). A down-vote penalizes the user receiving it by $-2$ points. Down-voting posts is costly ($-1$ point) in order to suppress trolling. Upon reaching certain milestones users can also earn reputational badges.

\subsection{Nestedness}

In ecological systems, nestedness refers to a property typically observed in the networks describing species-species interactions. Let us assume that such interactions in a given system are represented by a weighted bipartite adjacency matrix $W$, whose entry $w_{ij}$ quantifies the strength of interaction between species $i$ and $j$. In a perfectly nested matrix, an arrangement of rows and columns can be found such that the set of links in each row $i$ (column $j$) contains the set of links in row $i+1$ (column $j+1$), and such that matrix entries satisfy $W_{ij} \leq \min(W_{i-1,j},W_{i,j-1})$. It can be shown that among all possible connected bipartite networks with a fixed number of nodes and links, the one yielding the highest spectral radius $\rho(W)$ corresponds to a perfectly nested matrix~\cite{bell2008graphs}, where the spectral radius is defined as the largest singular value. Therefore, an ideal measure of nestedness in an empirical bipartite weighted matrix would be the ratio between its spectral ratio and that of the corresponding perfectly nested matrix with the same number of nodes and links. This, however, is unfeasible in practice due to the prohibitively high computational cost of identifying the perfectly nested matrix in the set via hard counting. Therefore, in our work we follow Staniczenko \emph{et al}.~\cite{staniczenko2013ghost}, and quantify the nestedness of a matrix with the $z$-score $z(\rho) = (\rho(W) - \overline{\rho}(W))/\sigma(\rho(W))$, where $\overline{\rho}(W)$ and $\sigma(\rho(W))$ represent, respectively, the mean and standard deviation of the spectral radii computed over a sampled population of bipartite matrices with the same nodes and edges as $W$, but with randomly reshuffled link weights.

\subsection{Logistic regression model for user specialization}

For each year in our analysis we calibrate the following logistic regression model:
\begin{equation} \label{eq:logistic_regression}
\log \left ( \frac{\pi_A}{1-\pi_A} \right ) = \beta_0^{A} +  \beta_{n}^{A} \log(n^A) + \beta_{t}^{A} \log(t^A) + \beta_H^{A} H^A + \beta_D^{A} D \ ,
\end{equation}
where $\pi_A$ denotes the probability that at least one of the answers posted by a user (with at least 10 posts in the year under consideration) gets accepted, i.e., marked as the best one in response to a question. In the above expression $n$ denotes the number of answers posted by a user, $t^A$ the number of tags associated with the corresponding questions, $H^A$ the specialization of the user as quantified by the Herfindahl index (see Eq.~\eqref{eq:herfindahl}), and $D$ the user's activity score (see Eq.~\eqref{eq:activity_score}).

\section*{References}
\vspace{0.01\textheight}
\bibliographystyle{naturemag}
\bibliography{lpm_bib}

\begin{thebibliography}{10}
\expandafter\ifx\csname url\endcsname\relax
  \def\url#1{\texttt{#1}}\fi
\expandafter\ifx\csname urlprefix\endcsname\relax\def\urlprefix{URL }\fi
\providecommand{\bibinfo}[2]{#2}
\providecommand{\eprint}[2][]{\url{#2}}

\bibitem{gottfried2016news}
\bibinfo{author}{Gottfried, J.} \& \bibinfo{author}{Shearer, E.}
\newblock \bibinfo{title}{News use across social media platforms 2016}
  (\bibinfo{year}{2016}).

\bibitem{zervas2017rise}
\bibinfo{author}{Zervas, G.}, \bibinfo{author}{Proserpio, D.} \&
  \bibinfo{author}{Byers, J.~W.}
\newblock \bibinfo{title}{The rise of the sharing economy: Estimating the
  impact of airbnb on the hotel industry}.
\newblock \emph{\bibinfo{journal}{Journal of marketing research}}
  \textbf{\bibinfo{volume}{54}}, \bibinfo{pages}{687--705}
  (\bibinfo{year}{2017}).

\bibitem{tadelis2015economics}
\bibinfo{author}{Tadelis, S.}
\newblock \bibinfo{title}{The economics of reputation and feedback systems in
  e-commerce marketplaces}.
\newblock \emph{\bibinfo{journal}{IEEE Internet Computing}}
  \textbf{\bibinfo{volume}{20}}, \bibinfo{pages}{12--19}
  (\bibinfo{year}{2015}).

\bibitem{tadelis2016reputation}
\bibinfo{author}{Tadelis, S.}
\newblock \bibinfo{title}{Reputation and feedback systems in online platform
  markets}.
\newblock \emph{\bibinfo{journal}{Annual Review of Economics}}
  \textbf{\bibinfo{volume}{8}}, \bibinfo{pages}{321--340}
  (\bibinfo{year}{2016}).

\bibitem{zloteanu2018digital}
\bibinfo{author}{Zloteanu, M.}, \bibinfo{author}{Harvey, N.},
  \bibinfo{author}{Tuckett, D.} \& \bibinfo{author}{Livan, G.}
\newblock \bibinfo{title}{Digital identity: The effect of trust and reputation
  information on user judgement in the sharing economy}.
\newblock \emph{\bibinfo{journal}{PloS one}} \textbf{\bibinfo{volume}{13}},
  \bibinfo{pages}{e0209071} (\bibinfo{year}{2018}).

\bibitem{livan2017excess}
\bibinfo{author}{Livan, G.}, \bibinfo{author}{Caccioli, F.} \&
  \bibinfo{author}{Aste, T.}
\newblock \bibinfo{title}{Excess reciprocity distorts reputation in online
  social networks}.
\newblock \emph{\bibinfo{journal}{Scientific reports}}
  \textbf{\bibinfo{volume}{7}}, \bibinfo{pages}{1--11} (\bibinfo{year}{2017}).

\bibitem{zervas2021first}
\bibinfo{author}{Zervas, G.}, \bibinfo{author}{Proserpio, D.} \&
  \bibinfo{author}{Byers, J.~W.}
\newblock \bibinfo{title}{A first look at online reputation on airbnb, where
  every stay is above average}.
\newblock \emph{\bibinfo{journal}{Marketing Letters}}
  \textbf{\bibinfo{volume}{32}}, \bibinfo{pages}{1--16} (\bibinfo{year}{2021}).

\bibitem{luca2016fake}
\bibinfo{author}{Luca, M.} \& \bibinfo{author}{Zervas, G.}
\newblock \bibinfo{title}{Fake it till you make it: Reputation, competition,
  and yelp review fraud}.
\newblock \emph{\bibinfo{journal}{Management Science}}
  \textbf{\bibinfo{volume}{62}}, \bibinfo{pages}{3412--3427}
  (\bibinfo{year}{2016}).

\bibitem{custodio2013generalists}
\bibinfo{author}{Cust{\'o}dio, C.}, \bibinfo{author}{Ferreira, M.~A.} \&
  \bibinfo{author}{Matos, P.}
\newblock \bibinfo{title}{Generalists versus specialists: Lifetime work
  experience and chief executive officer pay}.
\newblock \emph{\bibinfo{journal}{Journal of Financial Economics}}
  \textbf{\bibinfo{volume}{108}}, \bibinfo{pages}{471--492}
  (\bibinfo{year}{2013}).

\bibitem{brockman2016determinants}
\bibinfo{author}{Brockman, P.}, \bibinfo{author}{Lee, H. S.~G.} \&
  \bibinfo{author}{Salas, J.~M.}
\newblock \bibinfo{title}{Determinants of ceo compensation:
  Generalist--specialist versus insider--outsider attributes}.
\newblock \emph{\bibinfo{journal}{Journal of Corporate Finance}}
  \textbf{\bibinfo{volume}{39}}, \bibinfo{pages}{53--77}
  (\bibinfo{year}{2016}).

\bibitem{chen2021generalist}
\bibinfo{author}{Chen, G.}, \bibinfo{author}{Huang, S.},
  \bibinfo{author}{Meyer-Doyle, P.} \& \bibinfo{author}{Mindruta, D.}
\newblock \bibinfo{title}{Generalist versus specialist ceos and acquisitions:
  Two-sided matching and the impact of ceo characteristics on firm outcomes}.
\newblock \emph{\bibinfo{journal}{Strategic Management Journal}}
  \textbf{\bibinfo{volume}{42}}, \bibinfo{pages}{1184--1214}
  (\bibinfo{year}{2021}).

\bibitem{lazear2012leadership}
\bibinfo{author}{Lazear, E.~P.}
\newblock \bibinfo{title}{Leadership: A personnel economics approach}.
\newblock \emph{\bibinfo{journal}{Labour Economics}}
  \textbf{\bibinfo{volume}{19}}, \bibinfo{pages}{92--101}
  (\bibinfo{year}{2012}).

\bibitem{waller2019generalists}
\bibinfo{author}{Waller, I.} \& \bibinfo{author}{Anderson, A.}
\newblock \bibinfo{title}{Generalists and specialists: Using community
  embeddings to quantify activity diversity in online platforms}.
\newblock In \emph{\bibinfo{booktitle}{The World Wide Web Conference}},
  \bibinfo{pages}{1954--1964} (\bibinfo{year}{2019}).

\bibitem{onoue2013study}
\bibinfo{author}{Onoue, S.}, \bibinfo{author}{Hata, H.} \&
  \bibinfo{author}{Matsumoto, K.-i.}
\newblock \bibinfo{title}{A study of the characteristics of developers'
  activities in github}.
\newblock In \emph{\bibinfo{booktitle}{2013 20th Asia-Pacific Software
  Engineering Conference (APSEC)}}, vol.~\bibinfo{volume}{2},
  \bibinfo{pages}{7--12} (\bibinfo{organization}{IEEE}, \bibinfo{year}{2013}).

\bibitem{jiang2021wide}
\bibinfo{author}{Jiang, J.}, \bibinfo{author}{Maldeniya, D.},
  \bibinfo{author}{Lerman, K.} \& \bibinfo{author}{Ferrara, E.}
\newblock \bibinfo{title}{The wide, the deep, and the maverick: Types of
  players in team-based online games}.
\newblock \emph{\bibinfo{journal}{Proceedings of the ACM on Human-Computer
  Interaction}} \textbf{\bibinfo{volume}{5}}, \bibinfo{pages}{1--26}
  (\bibinfo{year}{2021}).

\bibitem{pike2019online}
\bibinfo{author}{Pike, C.~W.}, \bibinfo{author}{Zillioux, J.} \&
  \bibinfo{author}{Rapp, D.}
\newblock \bibinfo{title}{Online ratings of urologists: comprehensive
  analysis}.
\newblock \emph{\bibinfo{journal}{Journal of medical Internet research}}
  \textbf{\bibinfo{volume}{21}}, \bibinfo{pages}{e12436}
  (\bibinfo{year}{2019}).

\bibitem{movshovitz2013analysis}
\bibinfo{author}{Movshovitz-Attias, D.}, \bibinfo{author}{Movshovitz-Attias,
  Y.}, \bibinfo{author}{Steenkiste, P.} \& \bibinfo{author}{Faloutsos, C.}
\newblock \bibinfo{title}{Analysis of the reputation system and user
  contributions on a question answering website: Stackoverflow}.
\newblock In \emph{\bibinfo{booktitle}{2013 IEEE/ACM International Conference
  on Advances in Social Networks Analysis and Mining (ASONAM 2013)}},
  \bibinfo{pages}{886--893} (\bibinfo{organization}{IEEE},
  \bibinfo{year}{2013}).

\bibitem{staniczenko2013ghost}
\bibinfo{author}{Staniczenko, P.~P.}, \bibinfo{author}{Kopp, J.~C.} \&
  \bibinfo{author}{Allesina, S.}
\newblock \bibinfo{title}{The ghost of nestedness in ecological networks}.
\newblock \emph{\bibinfo{journal}{Nature communications}}
  \textbf{\bibinfo{volume}{4}}, \bibinfo{pages}{1--6} (\bibinfo{year}{2013}).

\bibitem{burghardt2017myopia}
\bibinfo{author}{Burghardt, K.}, \bibinfo{author}{Alsina, E.~F.},
  \bibinfo{author}{Girvan, M.}, \bibinfo{author}{Rand, W.} \&
  \bibinfo{author}{Lerman, K.}
\newblock \bibinfo{title}{The myopia of crowds: Cognitive load and collective
  evaluation of answers on stack exchange}.
\newblock \emph{\bibinfo{journal}{PloS one}} \textbf{\bibinfo{volume}{12}},
  \bibinfo{pages}{e0173610} (\bibinfo{year}{2017}).

\bibitem{bell2008graphs}
\bibinfo{author}{Bell, F.~K.}, \bibinfo{author}{Cvetkovi{\'c}, D.},
  \bibinfo{author}{Rowlinson, P.} \& \bibinfo{author}{Simi{\'c}, S.~K.}
\newblock \bibinfo{title}{Graphs for which the least eigenvalue is minimal, i}.
\newblock \emph{\bibinfo{journal}{Linear Algebra and its Applications}}
  \textbf{\bibinfo{volume}{429}}, \bibinfo{pages}{234--241}
  (\bibinfo{year}{2008}).

\end{thebibliography}

\section*{Acknowledgements}
G.L. acknowledges support from an EPSRC Early Career Fellowship (Grant No. EP/N006062/1).

\section*{Data availability}
The Stack Exchange data used in this paper are publicly accessible and can be downloaded via \href{https://archive.org/details/stackexchange}{https://archive.org/details/stackexchange}.

\appendix

\section{Additional tables and figures}

\begin{table}[htbp]
\centering
\caption{\textbf{Logistic regression model for the probability that a user's answer gets accepted}. We calibrate the model $\log(\pi_A / (1 - \pi_A)) = \beta_0^{A} + \beta_n^{A} \log(n^A) + \beta_t^{A} \log(t^A) + \beta_H^{A} H^A + \beta_D^{A} D$, where $\pi_A$ denotes the probability that a user has at least one accepted answer in a given year, $n^A$ indicates the number of answers posted by a user, $t^A$ the number of tags featured in the corresponding questions, $H^A$ the user's Herfindahl index (Eq.~\eqref{eq:herfindahl}), and $D$ the user's activity score (Eq.~\eqref{eq:activity_score}). The three bottom rows report -- respectively -- the number $N$ of users included in the model (i.e., users with at least 10 posts in the year of interest, of which at least one is an answer), the resulting regression model's area under the curve (AUC) and the model's $\chi^2$ statistic. The corresponding ROC curves are shown in \ref{fig:ROC}. Numbers in brackets indicate the standard errors of the estimated coefficients}
\begin{adjustbox}{width=\columnwidth,center}
\begin{tabular}{l c c c c c c c c c c c}
\hline \hline
					& 2009 	 & 2010 	 & 2011     & 2012 	  & 2013	  & 2014	  & 2015	  & 2016   	  & 2017     & 2018 	  & 2019           	\\ 
\hline
$\beta_0^A$ 		 & -4.29*** & -4.86*** & 4.79*** & -4.66***   & -4,39***  & -4.10*** & -4.15*** & -4.10*** & -3.93*** & -3.91*** & -3.77***	\\
            	      			 & (0.11)    & (0.09)    & (0.07)   & (0.06)      & (0.05)     & (0.04)	  & (0.04)	   & (0.04)	   & (0.04)	   & (0.04)	   & (0.04)		\\
$\beta_n^A$			 & 1.95***  & 1.92***  & 2.00*** & 1.95***     & 1.90***  & 2.02***  & 2.02***    & 2.03***  & 2.07***  & 2.28***  & 2.44***       	\\
            				 &  (0.09)   & (0.06)    & (0.05)   & (0.04)   	  & (0.03)	  & (0.03) 	  & (0.03)     & (0.03)	   & (0.03) 	   & (0.03) 	   & (0.03)		\\
$\beta_t^A$ 		 	 & 0.00      & 0.34***	 & 0.25*** & 0.22*** 	  & 0.07*     & -0.18*** & -0.20***  & -0.25*** & -0.38*** & -0.64*** & -0.91***		\\
            				 & (0.11) 	 & (0.08)	 & (0.07)	& (0.05)      & (0.04)	  & (0.04)    & (0.04)	   & (0.04) 	   & (0.04)    & (0.04) 	   & (0.04)		\\
$\beta_H^A$ 		 & 2.98*** 	 & 3.67***	 & 3.45*** & 3.37***     & 3.04***  & 2.73***  & 2.76***	   & 2.81***  & 2.59***  & 2.48***   & 2.23***    	\\
            				 & (0.19) 	 & (0.14)	 & (0.11)   & (0.09)      & (0.07)     & (0.07)    & (0.07) 	   & (0.07)	   & (0.07)	   & (0.07)    & (0.07) 	        \\		 
$\beta_D^A$ 		 &  -0.75***& -0.94***	 & -0.96***	& -0.96***   & -0.89***  & -0.81*** & -0.78*** & -0.74*** & -0.77*** & -0.80*** & -0.77***		 \\
            				 & (0.03)  	 & (0.03)	 & 0.02     & (0.02)      & (0.01)     & (0.12)	   & (0.02) 	   & (0.01)	   & (0.02) 	   & (0.01)	   & (0.01)		\\			
\hline
$N$           			& 33,669    & 60,035  & 100,568 & 151,132 & 198,553 & 221,941 & 237,108 & 241,120 & 232,722   & 207,945	  & 199,795	\\          
$\mathrm{AUC}$           	& 0.84	 & 0.83    	 & 0.83	 & 0.82	  & 0.80	  & 0.79	  & 0.78	  & 0.78	   & 0.77	  & 0.76	  & 0.76	\\
$\chi^2$ statistics ($\times 10^4$)& 1.01***    & 1.85*** & 3.07*** & 4.36*** & 5.52** & 5.87*** & 6.31*** & 6.23*** & 5.69***  & 5.10*** & 4.96*** \\
\hline \hline
\multicolumn{3}{l}{\textsuperscript{***}$p<0.01$, 
  \textsuperscript{**}$p<0.05$, 
  \textsuperscript{*}$p<0.1$}
      \label{tab:regression_acceptance}
\end{tabular}
\end{adjustbox}
\end{table}

\begin{table}[htbp]
\centering
\caption{\textbf{Multinomial logistic regression model for the probability that a user's answers attract votes.} We calibrate the model
$\log (\pi^{(u,A} / \pi^{(z,A)})  = \beta_0^{(u,A)} + \beta_{n}^{(u,A)} \log(n^A) + \beta_{t}^{(u,A)} \log(t^A) + \beta_H^{(u,A)} H^A + \beta_D^{(u,A)} D$; $\log (\pi^{(u,A} / \pi^{(z,A)})  = \beta_0^{(v,A)} + \beta_{n}^{(v,A)} \log(n^A) + \beta_{t}^{(v,A)} \log(t^A) + \beta_H^{(v,A)} H^A + \beta_D^{(v,A)} D$, where $\pi^{(z,A)}$, $\pi^{(u,A)}$ and $\pi^{(v,A)}$ indicate -- respectively -- the probability that a user's posted answers receive zero votes, only up-votes, and both up- and down-votes in a given year. $n^A$ indicates the number of answers posted by a user, $t^A$ the number of tags featured in the corresponding questions, $H^A$ the user's Herfindahl index (Eq.~\eqref{eq:herfindahl}), and $D$ the user's activity score (Eq.~\eqref{eq:activity_score}). The four bottom rows report the total number of users ($N$), and the fractions of users whose posted answers received only up-votes ($N_u$), both up- and down-votes ($N_v$) and zero votes ($N_z$). Numbers in brackets indicate the standard errors of the estimated coefficients.}
\begin{adjustbox}{width=\columnwidth,center}
\begin{tabular}{l c c c c c c c c c c c}
\hline \hline
			 	& 2009 	 & 2010 	 & 2011     & 2012 	  & 2013	  & 2014	  & 2015	  & 2016   	  & 2017     & 2018 	 & 2019           	\\ 
\hline
$\beta_0^{(u,A)}$      & -2.43***   & -2.54*** & -2.64*** & -2.72***  & -2.84*** & -2.99** & -3.12*** & -3.35***  & -3.51*** & -3.79*** & -3.78***	\\
            	      		& (0.09)      & (0.07)    & (0.06)    & (0.06)     & (0.05)    & (0.06)	  & (0.06)	  & (0.06)	  & (0.07)	  & (0.08)	  & (0.10)		\\
$\beta_n^{(u,A)}$   	&  -0.05      & -0.11     & -0.12**  &  -0.03      & -0.03      & 0.00     & 0.02      & -0.16***  & -0.04     & -0.06     & 0.00      	\\
            			&  (0.10)    & (0.07)   	 & (0.06)    & (0.05)   	   & (0.04)	   & (0.04) 	  & (0.04)    & (0.05)	  & (0.05) 	  & (0.06) 	  & (0.07)		\\
$\beta_t^{(u,A)}$   	&  0.24*     & 0.26***  & 0.27***  &  0.12*      & 0.12**    & 0.08     & 0.05       & 0.26***  & 0.12*     & 0.18**    & 0.02       	\\
            			&  (0.12)    & (0.09)   	 & (0.08)    & (0.06)   	  & (0.06)	   & (0.06) 	  & (0.06)    & (0.07)	  & (0.07) 	  & (0.08) 	  & (0.09)		\\			
$\beta_H^{(u,A)}$     	&  0.40*     & 0.44***   & 0.09     & 0.19* 	  & 0.10       & 0.08	  & 0.04      & 0.24*	  & 0.15	  & 0.45***  & 0.13		\\
            			&  (0.22)    & (0.16)    & (0.14)    & (0.12)     & (0.11)	   & (0.12)	  & (0.12)    & (0.13)	  & (0.15)    & (0.16)    & (0.18)		\\
$\beta_D^{(u,A)}$     &  -0.10**	 & 0.06*     & -0.03     & 0.03	  & -0.01      & 0.00	  & 0.00	  & 0.06***	  & 0.00      & 0.02  	  & 0.06*		\\
            			&  (0.04)   	 & (0.03)    & (0.02)    & (0.02)  	  & (0.02)	   & (0.02)	  & (0.02)	  & (0.02) 	  & (0.03)	  & (0.03) 	  & (0.03)		\\
\hline
$\beta_0^{(v,A)}$      & -2.15***  & -2.97*** & -3.28*** & -3.45***  & -3.49*** & -4.01*** & -3.98*** & -4.47*** & -4.73*** & -5.12*** & -5.61***	\\
            	      		& (0.09)     & (0.08)    & (0.07)    & (0.08)     & (0.08)    & (0.09)	  & (0.10)	  & (0.08)	  & (0.13)	  & (0.16)	  & (0.20)		\\
$\beta_n^{(v,A)}$   	&  0.07      & -0.09      & -0.14*   & 0.00        & 0.05	  & -0.04     & 0.14* 	  & -0.01 	  & -0.09      & 0.36*** & -0.07       	\\
            			&  (0.09)    & (0.08)   	 & (0.07)    & (0.07)   	  & (0.06)	  & (0.07) 	  & (0.07)    & (0.08)	  & (0.09) 	  & (0.11) 	  & (0.14)		\\
$\beta_t^{(v,A)}$   	&  0.10      & 0.31***   & 0..34*** & 0.13       & -0.04	  & 0.16*     & -0.14 	  & 0.08 	  & 0.16      & 0.29**    & 0.18       	\\
            			&  (0.10)    & (0.10)   	 & (0.09)    & (0.09)   	  & (0.09)	  & (0.09) 	  & (0.10)    & (0.11)	  & (0.13) 	  & (0.15) 	  & (0.20)		\\			
$\beta_H^{(v,A)}$     	&  -0.20     & 0.55***  & 0.51***   & 0.19 	  & 0.02      & 0.08	  & -0.33  	  & 0.27 	  & 0.40   	  & 0.20      & 0.68*		\\
            			&  (0.22)    & (0.18)    & (0.16)    & (0.16)     & (0.16)	 & (0.19)	  & (0.21)    & (0.22)	  & (0.25)    & (0.31)    & (0.36)		\\
$\beta_D^{(v,A)}$      &  -0.11*** & 0.00 	 & -0.03     & -0.03	  & 0.04      & -0.03	  & -0.07**	  & 0.03	  & 0.03  	  & -0.15*    & 0.06		\\
            			&  (0.04)   	 & (0.03)    & (0.03)    & (0.03)  	  & (0.03)	  & (0.03)	  & (0.03)	  & (0.04) 	  & (0.05)	  & (0.06) 	  & (0.07)		\\
\hline		
$N$           		& 33,669    & 60,035  & 100,573 & 151,132 & 198,553 & 221,941& 237,108 & 241,120 & 232,722 & 207,950 & 199,795	\\          
$N_u (\%)$           	& 11.3 \%	 & 10.1\%  & 8.69\%	  & 7.48\%	  & 6.48\%	  & 5.57\%	  & 4.89\%	   & 4.21\%   & 3.51\%   & 2.97\%  & 2.41\%	\\
$N_v (\%)$		& 13.7\%	 & 7.90\%	 & 5.59\%   & 3.94\%	  & 2.91\%	  & 2.15\%	  & 1.72\%	  & 1.32\%	   & 1.05\%   & 0.76\%  & 0.51\%      \\
$N_z (\%)$		& 74.9\%   & 81.9\%  & 85.6\%   & 88.5\%   & 90.5\%  & 92.2\%  & 93.3\%  & 94.3\%    & 95.4\%   & 96.2\%  & 97.0\%     \\ 
\hline \hline
\multicolumn{3}{l}{\textsuperscript{***}$p<0.01$, 
  \textsuperscript{**}$p<0.05$, 
  \textsuperscript{*}$p<0.1$}
  \label{tab:multinomial_A}
\end{tabular}
\end{adjustbox}
\end{table}

\begin{table}[htbp]
\centering
\caption{\textbf{Multinomial logistic regression model for the probability that a user's questions attract votes.} We calibrate the model
$\log (\pi^{(u,Q} / \pi^{(z,Q)})  = \beta_0^{(u,Q)} + \beta_{n}^{(u,Q)} \log(n^Q) + \beta_{t}^{(u,Q)} \log(t^Q) + \beta_H^{(u,Q)} H^Q + \beta_D^{(u,Q)} D$; $\log (\pi^{(u,Q} / \pi^{(z,Q)})  = \beta_0^{(v,Q)} + \beta_{n}^{(v,Q)} \log(n^Q) + \beta_{t}^{(v,Q)} \log(t^Q) + \beta_H^{(v,Q)} H^Q + \beta_D^{(v,Q)} D$, where $\pi^{(z,Q)}$, $\pi^{(u,Q)}$ and $\pi^{(v,Q)}$ indicate -- respectively -- the probability that a user's posted questions receive zero votes, only up-votes, and both up- and down-votes in a given year. $n^Q$ indicates the number of questions posted by a user, $t^Q$ the number of tags featured in the such questions, $H^Q$ the user's Herfindahl index (Eq.~\eqref{eq:herfindahl}), and $D$ the user's activity score (Eq.~\eqref{eq:activity_score}). The four bottom rows report the total number of users ($N$), and the fractions of users whose posted quesions received only up-votes ($N_u$), both up- and down-votes ($N_v$) and zero votes ($N_z$). Numbers in brackets indicate the standard errors of the estimated coefficients.}
\begin{adjustbox}{width=\columnwidth,center}
\begin{tabular}{l c c c c c c c c c c c}
\hline \hline
				& 2009 	 & 2010 	 & 2011      & 2012 	   & 2013	  & 2014	  & 2015	  & 2016   	  & 2017     & 2018 	 & 2019           	\\ 
\hline
$\beta_0^{(u,Q)}$      & -1.88***   & -1.85*** & -2.38*** & -2.53***  & -2.71*** & -3.02*** & -3.02***& -3.37***  & -3.58*** & -3.72*** & -3.89***	\\
            	      		& (0.11)      & (0.08)    & (0.07)    & (0.06)     & (0.06)    & (0.12)	   & (0.08)	  & (0.08)	  & (0.10)	  & (0.11)	  & (0.13)		\\
$\beta_n^{(u,Q)}$   	&  0.00       & 0.11**    & 0.01       &  0.04      & -0.08**  & -0.05      & 0.01     & -0.05      & -0.02     & -0.07	  & -0.06       	\\
            			&  (0.07)    & (0.05)   	  & (0.04)    & (0.03)     & (0.03)	   & (0.04) 	   & (0.04)   & (0.04)	  & (0.05) 	  & (0.05) 	  & (0.06)		\\
$\beta_t^{(u,Q)}$   	&  0.15       & -0.08     & 0.05       &  0.00      & 0.12**    & 0.09*     & -0.03    & 0.10*      & 0.06      & 0.10	  & 0.00       	\\
            			&  (0.10)    & (0.07)   	  & (0.06)    & (0.05)     & (0.05)	   & (0.05) 	   & (0.05)   & (0.06)	  & (0.07) 	  & (0.08) 	  & (0.09)		\\			
$\beta_H^{(u,Q)}$     &  -0.20     & 0.20        & 0.00      & 0.09 	   & 0.01      & 0.29**	   & 0.00     & 0.19	  & 0.02	  & 0.08      & 0.09		\\
            			&  (0.22)    & (0.16)     & (0.14)    & (0.13)     & (0.13)	   & (0.13)	   & (0.14)   & (0.15)	  & (0.18)    & (0.21)    & (0.23)		\\
$\beta_D^{(u,Q)}$      &  0.09***	 & 0.08***   & 0.04**    & 0.03	   & 0.07***  & 0.01	   & 0.03	  & 0.03	  & 0.03      & 0.05*  	  & -0.03		\\
            			&  (0.03)   	 & (0.02)     & (0.02)    & (0.02)  	   & (0.02)	   & (0.02)	   & (0.02)	  & (0.02) 	  & (0.02)	  & (0.03) 	  & (0.03)		\\
\hline
$\beta_0^{(v,Q)}$      & -2.90***  & -3.07*** & -3.09*** & -3.37***   & -3.60*** & -3.60*** & -3.82*** & -4.19*** & -4.29*** & -4.77***& -4.85***	\\
            	      		& (0.15)     & (0.14)    & (0.13)    & (0.11)      & (0.10)    & (0.12)	   & (0.12)	  & (0.14)	  & (0.16)	  & (0.18)	  & (0.21)		\\
$\beta_n^{(v,Q)}$   	&  -0.07     & 0.02       & 0.14**   & 0.00        & 0.00      & 0.01       & 0.03      & 0.02 	  & 0.03      & 0.06       & 0.00       	\\
            			&  (0.11)    & (0.09)   	 & (0.07)    & (0.06)   	  & (0.05)	  & (0.06) 	   & (0.06)    & (0.07)	  & (0.07) 	  & (0.09) 	  & (0.10)		\\			
$\beta_t^{(v,Q)}$   	&  0.27*     & -0.03     & -0.26*** & -0.07       & -0.09     & -0.17*    & -0.15*   & -0.08 	  & -0.13     & -0.04      & 0.00       	\\
            			&  (0.14)    & (0.12)   	 & (0.10)    & (0.08)   	  & (0.08)	  & (0.09) 	  & (0.09)    & (0.10)	  & (0.11) 	  & (0.13) 	  & (0.14)		\\
$\beta_H^{(v,Q)}$     	&  0.29     	 & 0.38      & -0.36     & 0.07 	  & 0.22      & -0.30 	  & -0.12  	  & 0.02 	  & -0.16   	  & 0.08      & -0.31		\\
            			&  (0.30)    & (0.27)    & (0.25)    & (0.21)     & (0.19)	  & (0.23)	  & (0.23)    & (0.25)	  & (0.29)    & (0.34)    & (0.40)		\\
$\beta_D^{(v,Q)}$      &  0.07*	 & 0.01 	 & 0.02      & -0.02	  & -0.02     & -0.03	  & 0.02	  & -0.01	  & 0.00  	  & 0.06      & 0.09		\\
            			&  (0.04)   	 & (0.04)    & (0.03)    & (0.03)  	  & (0.03)	  & (0.03)	  & (0.03)	  & (0.04) 	  & (0.04)	  & (0.05) 	  & (0.05)		\\
\hline			
$N$           		& 30,421    & 59,241  & 104,794 & 158,229 & 209,674 & 234,670& 246,786 & 246,786 & 240,697 & 213,597 & 214,424	\\          
$N_u (\%)$           	& 16.3 \%	 & 14.5\%  & 10.6\%	  & 8.01\%	  & 6.41\%	  & 5.11\%	  & 4.41\%	   & 3.62\%   & 2.92\%   & 2.43\%  & 1.82\%	\\
$N_v (\%)$		& 7.03\%	 & 3.96\%	 & 2.92\%   & 2.64\%	  & 2.19\%	  & 1.72\%	  & 1.57\%	  & 1.28\%	   & 1.05\%   & 0.87\%  & 0.69\%      \\
$N_z (\%)$		& 76.7\%   & 81.5\%  & 86.5\%   & 89.3\%   & 91.3\%  & 93.0\%  & 93.9\%  & 95.0\%    & 95.9\%   & 96.6\%  & 97.4\%     \\ 
\hline \hline
\multicolumn{3}{l}{\textsuperscript{***}$p<0.01$, 
  \textsuperscript{**}$p<0.05$, 
  \textsuperscript{*}$p<0.1$}
  \label{tab:multinomial_Q}
\end{tabular}
\end{adjustbox}
\end{table}

\begin{table}[htbp]
\centering
\caption{\textbf{Stepwise linear regression model for the number of up-votes received by a user's answers}. We calibrate the model $\log (v^{\uparrow A})  = \beta_0^{\uparrow A} + \beta_{n}^{\uparrow A} \log(n^A) + \beta_{t}^{\uparrow A} \log(t^A) + \beta_H^{\uparrow A} H^A + \beta_D^{\uparrow A} D$, where $v^{\uparrow A}$ denotes the number of up-votes received by a user's posted answers, $n^A$ indicates the number of answers posted by a user, $t^A$ the number of tags featured in the corresponding questions, $H^A$ the user's Herfindahl index (Eq.~\eqref{eq:herfindahl}), and $D$ the user's activity score (Eq.~\eqref{eq:activity_score}). The three bottom rows report -- respectively -- the number $N$ of users included in the model (i.e., users with at least 10 posts in the year of interest and with at least one up-vote to their posted answers), the resulting regression model's $R^2$ coefficient and the model's $F$ statistic. Numbers in brackets indicate the standard errors of the estimated coefficients.}
\begin{adjustbox}{width=\columnwidth,center}
\begin{tabular}{l c c c c c c c c c c c}
\hline \hline
\textbf{Variable} 		& 2009 	 & 2010 	 & 2011     & 2012 	  & 2013	  & 2014	  & 2015	  & 2016   	  & 2017     & 2018 	 & 2019           	\\ 
\hline
$\beta_0^{\uparrow A}$ 	& 2.96***   & 2.61***  & 1.90***  & 2.35***   & 2.37***   & 2.03***  & 1.96***   & 1.88***  & 1.69***  & 1.75***  & 1.27***	\\
            	      			& (0.06)     & (0.04)    & (0.20)    & (0.04)     & (0.04)    & (0.03)	  & (0.03)	  & (0.03)	  & (0.03)	  & (0.07)	  & (0.04)		\\
$\beta_n^{\uparrow A}$ 	& -0.91***  & -0.91*** & -1.07*** & -1.08***  & -1.01*** & -0.96*** & -0.98*** & -0.98*** & -0.98*** & -0.93*** & -0.98***       	\\
            				&  (0.02)    & (0.01)   	 & (0.08)    & (0.06)   	  & (0.01)	  & (0.01) 	  & (0.01)    & (0.01)	  & (0.01) 	  & (0.07) 	  & (0.02)		\\
$\beta_t^{\uparrow A}$ 	&  -        	 & -		 & 0.44***	 & 0.17**     & -		  & -		  & -		  & -		  & -		  & -0.20**	  & -		\\
            				&  	    	 & 	         & (0.13)	 & (0.08)     & 		  &		  &		  & 		  &		  & (0.08)	  & 		\\
$\beta_H^{\uparrow A}$	&  -	 	 & -	         & 1.22***	 & -		  & -		  & -		  & -		  & -		  & -		  & -		  & -    	\\
            				&  	   	 & 	         & (0.01)	 &		  & 		  &		  & 		  & 		  & 		  & -	 	  & 	        \\		 
$\beta_D^{\uparrow A}$	&  - 		 & -	         & -		 & -		  & 0.08***  & -		  & -		  & -		  & -		  & -		  & -		\\
            				&  	   	 & 	         & 		 &		  & (0.03)	  & 		  &  		  & 		  & 		  &		  & 		\\			
\hline
$N$           			& 8,423    	 & 10,787 	 & 14,364   & 17,257	  & 18,643	  & 17,134	  & 15,683	  & 13,318	   & 10,610  & 7,775	  & 5,842	\\          
$R^{2}$           			& 0.28	 & 0.31    	 & 0.33	 & 0.33	  & 0.35	  & 0.34	  & 0.35	  & 0.36	   & 0.38	  & 0.41	  & 0.42	\\
$F$ statistic ($\times 10^3$)& 3.35***    & 4.90*** & 1.78*** & 4.22*** & 4.96*** & 8.83***	  & 8.56***	   & 7.49*** & 6.36*** & 1.79*** & 4.18*** \\
\hline \hline
\multicolumn{3}{l}{\textsuperscript{***}$p<0.01$, 
  \textsuperscript{**}$p<0.05$, 
  \textsuperscript{*}$p<0.1$}
    \label{tab:stepwise_A_up}
\end{tabular}
\end{adjustbox}
\end{table}

\begin{table}[htbp]
\centering
\caption{\textbf{Stepwise linear regression model for the number of down-votes received by a user's answers}. We calibrate the model $\log (v^{\downarrow A})  = \beta_0^{\downarrow A} + \beta_{n}^{\downarrow A} \log(n^A) + \beta_{t}^{\downarrow A} \log(t^A) + \beta_H^{\downarrow A} H^A + \beta_D^{\downarrow A} D$, where $v^{\downarrow A}$ denotes the number of down-votes received by a user's posted answers, $n^A$ indicates the number of answers posted by a user, $t^A$ the number of tags featured in the corresponding questions, $H^A$ the user's Herfindahl index (Eq.~\eqref{eq:herfindahl}), and $D$ the user's activity score (Eq.~\eqref{eq:activity_score}). The three bottom rows report -- respectively -- the number $N$ of users included in the model (i.e., users with at least 10 posts in the year of interest and with at least one down-vote to their posted answers), the resulting regression model's $R^2$ coefficient and the model's $F$ statistic. Numbers in brackets indicate the standard errors of the estimated coefficients.}
\begin{adjustbox}{width=\columnwidth,center}
\begin{tabular}{l c c c c c c c c c c c}
\hline \hline
\textbf{Variable} 		& 2009 	 & 2010 	 & 2011     & 2012 	  & 2013	  & 2014	  & 2015	  & 2016   	  & 2017     & 2018 	 & 2019           	\\ 
\hline
$\beta_0^{\downarrow A}$ & 1.27***   & 1.03***  & 0.90***  & 0.93***   & 0.72***   & 0.61***  & 0.58***  & 0.59***  & 0.49***  & 0.45***   & 0.36***	\\
            	      			 & (0.05)     & (0.04)    & (0.06)   & (0.05)     & (0.03)    & (0.03)	  & (0.03)	  & (0.03)	  & (0.04)	  & (0.04)	  & (0.07)		\\
$\beta_n^{\downarrow A}$ & -0.97***  & -0.97*** & -0.98*** & -1.02***  & -1.00*** & -0.99*** & -1.00*** & -1.01*** & -0.98*** & -0.98*** & -0.93***       	\\
            				 &  (0.01)    & (0.01)   & (0.01)    & (0.01)   	  & (0.01)	  & (0.01) 	  & (0.01)    & (0.01)	  & (0.01) 	  & (0.01) 	  & (0.03)		\\
$\beta_t^{\downarrow A}$ 	 &  -        	 & -		 & -		 & - 	          & -		  & -		  & -		  & -		  & -		  & -		  & -		\\
            				 &  	    	 & 	         &		 &               & 		  &		  &		  & 		  &		  & 		  & 		\\
$\beta_H^{\downarrow A}$ &  -	 	 & -	         & 0.27**	 & -0.27**	  & -		  & -		  & -		  & -		  & -		  &     	  & -    	\\
            				 &  	   	 & 	         & (0.13)	 & (0.12)	  & 		  &		  & 		  & 		  & 		  &     	  & 	        \\		 
$\beta_D^{\downarrow A}$ &  - 		 & -	         & -		 & -		  & 	          & -		  & -		  & -		  & -		  & -		  & -0.07**	 \\
            				 &  	   	 & 	         & 		 &		  & 		  & 		  &  		  & 		  & 		  &		  & (0.03)	\\			
\hline
$N$           			& 4,659    	 & 4,816	 & 5,721   & 6,080	  & 5,937	  & 4,973	  & 4,302	  & 3,379	   & 2,606  & 1,755	  & 1,131	\\          
$R^{2}$           			& 0.55	 & 0.62    	 & 0.64	 & 0.67	  & 0.67	  & 0.68	  & 0.69	  & 0.69	   & 0.69	  & 0.71	  & 0.71	\\
$F$ statistisc ($\times 10^3$)& 5.58***    & 7.78*** & 5.03*** & 6.25*** & 10.6*** & 10.6***	  & 9.70***	   & 7.43*** & 5.80*** & 2.72*** & 0.94*** \\
\hline \hline
\multicolumn{3}{l}{\textsuperscript{***}$p<0.01$, 
  \textsuperscript{**}$p<0.05$, 
  \textsuperscript{*}$p<0.1$}
      \label{tab:stepwise_A_down}
\end{tabular}
\end{adjustbox}
\end{table}

\begin{table}[htbp]
\centering
\caption{\textbf{Stepwise linear regression model for the number of up-votes received by a user's questions}. We calibrate the model $\log (v^{\uparrow Q})  = \beta_0^{\uparrow Q} + \beta_{n}^{\uparrow Q} \log(n^Q) + \beta_{t}^{\uparrow Q} \log(t^Q) + \beta_H^{\uparrow Q} H^Q + \beta_D^{\uparrow Q} D$, where $v^{\uparrow Q}$ denotes the number of up-votes received by a user's posted answers, $n^Q$ indicates the number of answers posted by a user, $t^Q$ the number of tags featured in the corresponding questions, $H^Q$ the user's Herfindahl index (Eq.~\eqref{eq:herfindahl}), and $D$ the user's activity score (Eq.~\eqref{eq:activity_score}). The three bottom rows report -- respectively -- the number $N$ of users included in the model (i.e., users with at least 10 posts in the year of interest and with at least one up-vote to their posted questions), the resulting regression model's $R^2$ coefficient and the model's $F$ statistic. Numbers in brackets indicate the standard errors of the estimated coefficients.}
\begin{adjustbox}{width=\columnwidth,center}
\begin{tabular}{l c c c c c c c c c c c}
\hline \hline
\textbf{Variable} 		& 2009 	 & 2010 	 & 2011     & 2012 	  & 2013	  & 2014	  & 2015	  & 2016   	  & 2017     & 2018 	 & 2019           	\\ 
\hline
$\beta_0^{\uparrow Q}$ 	& 2.72***   & 2.28***  & 2.28***  & 2.11***   & 1.80***   & 1.47***  & 1.49***   & 1.39***  & 1.28***  & 1.01***   & 0.86***	\\
            	      			& (0.06)     & (0.05)    & (0.04)    & (0.06)     & (0.03)    & (0.03)	  & (0.03)	  & (0.03)	  & (0.04)	  & (0.04)	  & (0.04)		\\
$\beta_n^{\uparrow Q}$ 	& -0.89***  & -1.17*** & -1.08*** & -1.13***  & -0.98*** & -1.01*** & -0.97*** & -0.96*** & -0.98*** & -0.95*** & -0.96***       	\\
            				&  (0.02)    & (0.07)   	 & (0.04)    & (0.04)   	  & (0.01)	  & (0.03) 	  & (0.01)    & (0.01)	  & (0.01) 	  & (0.01) 	  & (0.02)		\\
$\beta_t^{\uparrow Q}$ 	&  -        	 & 0.28***	 & 0.11**	 & -             & -		  & 0.08**   & -		  & -		  & -		  & -		  & -		\\
            				&  	    	 & (0.07)	 & (0.05)    & 	          & 		  & (0.04)	  & 	          & 		  &		  & 		  & 		\\
$\beta_H^{\uparrow Q}$	&  -	 	 & -	         & -		 & -		  & -		  & -		  & -		  & -		  & -		  & -		  & -    	\\
            				&  	   	 & 	         &		 &		  & 		  &		  & 		  & 		  & 		  & 	 	  & 	        \\		 
$\beta_D^{\uparrow Q}$	&  0.12 	 & -	         & -		 & -		  &  	          & -		  & 0.05**	  & -		  & -		  & -		  & -		\\
            				&  (0.04)	 & 	         & 		 &		  & 		  & 		  & (0.02) 	  & 		  & 		  &		  & 		\\			
\hline
$N$           			& 7,089    	 & 10,923 	 & 14,129   & 16,850	  & 18,033	  & 16,041	  & 14,763  & 12,175	   & 9,564  & 7,049	  & 5,377	\\          
$R^{2}$           			& 0.23	 & 0.26    	 & 0.29	 & 0.30	  & 0.31	  & 0.30	  & 0.32	  & 0.33	   & 0.35	  & 0.37	  & 0.41	\\
$F$ statistisc ($\times 10^3$)& 1.05***    & 1.90*** & 2.92*** & 2.42*** & 8.07*** & 3.48***	  & 3.52***	   & 5.95*** & 5.12*** & 3.71*** & 3.71*** \\
\hline \hline
\multicolumn{3}{l}{\textsuperscript{***}$p<0.01$, 
  \textsuperscript{**}$p<0.05$, 
  \textsuperscript{*}$p<0.1$}
      \label{tab:stepwise_Q_up}
\end{tabular}
\end{adjustbox}
\end{table}

\begin{table}[htbp]
\centering
\caption{\textbf{Stepwise linear regression model for the number of down-votes received by a user's questions}. We calibrate the model $\log (v^{\downarrow Q})  = \beta_0^{\downarrow Q} + \beta_{n}^{\downarrow Q} \log(n^Q) + \beta_{t}^{\downarrow Q} \log(t^Q) + \beta_H^{\downarrow Q} H^Q + \beta_D^{\downarrow Q} D$, where $v^{\downarrow Q}$ denotes the number of down-votes received by a user's posted answers, $n^Q$ indicates the number of answers posted by a user, $t^Q$ the number of tags featured in the corresponding questions, $H^Q$ the user's Herfindahl index (Eq.~\eqref{eq:herfindahl}), and $D$ the user's activity score (Eq.~\eqref{eq:activity_score}). The three bottom rows report -- respectively -- the number $N$ of users included in the model (i.e., users with at least 10 posts in the year of interest and with at least one down-vote to their posted questions), the resulting regression model's $R^2$ coefficient and the model's $F$ statistic. Numbers in brackets indicate the standard errors of the estimated coefficients.}
\begin{adjustbox}{width=\columnwidth,center}
\begin{tabular}{l c c c c c c c c c c c}
\hline \hline
\textbf{Variable} 		& 2009 	 & 2010 	 & 2011     & 2012 	  & 2013	  & 2014	  & 2015	  & 2016   	  & 2017     & 2018 	 & 2019           	\\ 
\hline
$\beta_0^{\downarrow Q}$ & 0.68***   & 0.61***  & 0.62*** & 0.66***   & 0.66***   & 0.59***  & 0.66***  & 0.54***  & 0.54***  & 0.54***   & 0.44***	\\
            	      			 & (0.05)     & (0.04)    & (0.04)   & (0.03)     & (0.03)    & (0.03)	  & (0.04)	  & (0.04)	  & (0.04)	  & (0.05)	  & (0.05)		\\
$\beta_n^{\downarrow Q}$ & -0.98*** & 1.02***  & -1.03*** & -1.03***  & -1.02*** & -1.01*** & -1.03*** & -0.99*** & -1.00*** & -1.01*** & -0.97***       	\\
            				 &  (0.02)   & (0.02)    & (0.01)    & (0.01)   	  & (0.01)	  & (0.01) 	  & (0.01)    & (0.01)	  & (0.02) 	  & (0.02) 	  & (0.02)		\\
$\beta_t^{\downarrow Q}$ &  -        	 & -		 & -		 & - 	          & -		  & -		  & -		  & -		  & -		  & -		  & -		\\
            				 &  	    	 & 	         &		 &               & 		  &		  &		  & 		  &		  & 		  & 		\\
$\beta_H^{\downarrow Q}$ &  -	 	 & -	         & -		 & -		  & -		  & -		  & -		  & -		  & -		  &     	  & -    	\\
            				 &  	   	 & 	         &		 & 		  & 		  &		  & 		  & 		  & 		  &     	  & 	        \\		 
$\beta_D^{\downarrow Q}$ &  - 	 & -	         & -		 & -		  & 	          & -		  & -		  & -		  & -		  & -		  & 		 \\
            				 &  	   	 & 	         & 		 &		  & 		  & 		  &  		  & 		  & 		  &		  & 		\\			
\hline
$N$           			& 2,144    	 & 2.387 	 & 3,130   & 4,300	  & 4,831	  & 4,358	  & 4,179	  & 3,485	   & 2,830   & 2,079	  & 1,718	\\          
$R^{2}$           			& 0.62	 & 0.65    	 & 0.66	 & 0.63	  & 0.62	  & 0.61	  & 0.59	  & 0.58	   & 0.58	  & 0.61	  & 0.58	\\
$F$ statistisc ($\times 10^3$)& 3.50***    & 4.51*** & 6.08*** & 7.84*** & 7.84*** & 6.83***	  & 6.11***	   & 3.89*** & 3.89***  & 3.21*** & 2.34*** \\
\hline \hline
\multicolumn{3}{l}{\textsuperscript{***}$p<0.01$, 
  \textsuperscript{**}$p<0.05$, 
  \textsuperscript{*}$p<0.1$}
      \label{tab:stepwise_Q_down}
\end{tabular}
\end{adjustbox}
\end{table}

\newpage

\begin{figure*}[h!]
\centering
\includegraphics[scale=0.5]{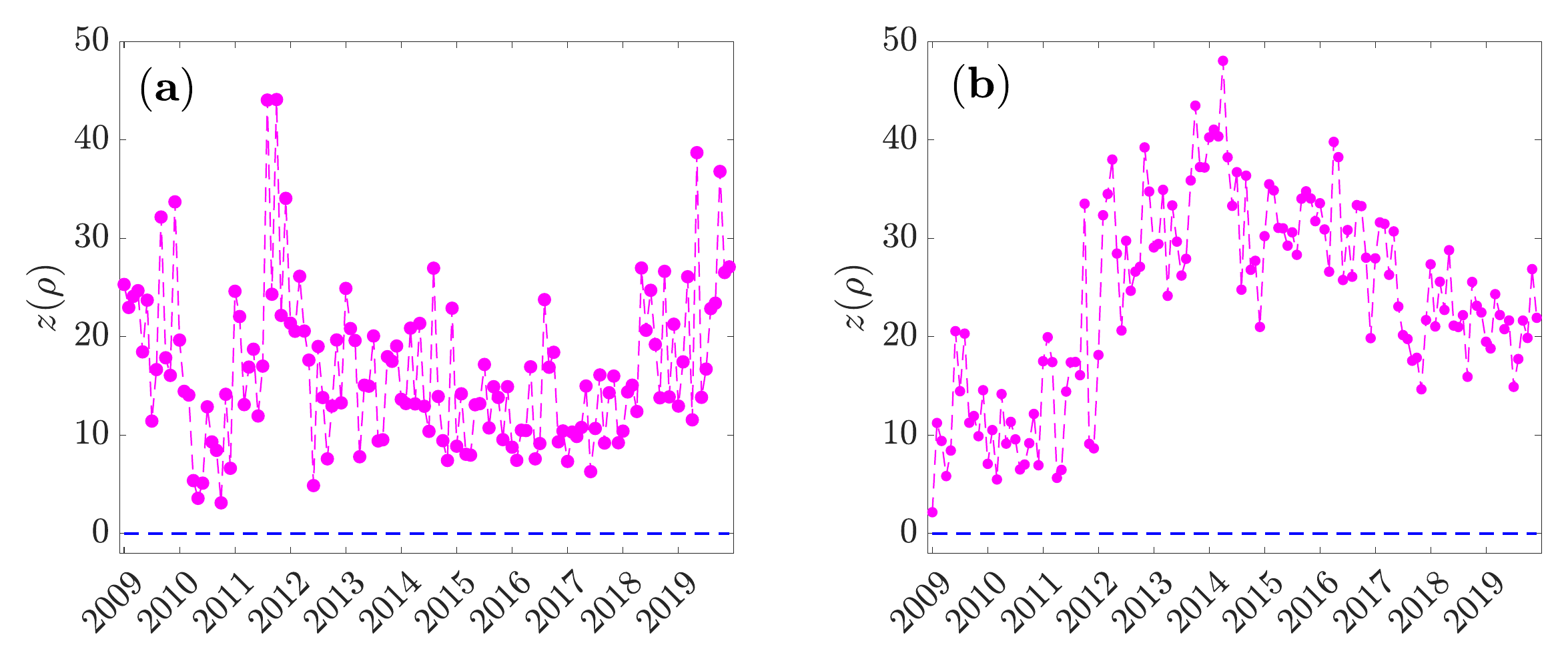}
\caption{\textbf{Evidence of nestedness in Stack Overflow's user-tag bipartite networks.} $\mathbf{a}$ $z$-scores of the spectral radius $\rho$ calculated in the monthly user-tag networks of $A$-users (i.e., users with activity score $D = 1$, see Eq.~\eqref{eq:activity_score}). $\mathbf{b}$ Same quantity calculated in the monthly user networks of $Q$-users ($D=-1$). In both panels, $z$-scores are calculated with the procedure put forward in~\cite{staniczenko2013ghost}.}
\label{fig:nestedness}
\end{figure*}

\begin{figure*}[h!]
	\centering
	\includegraphics[width=.85\linewidth]{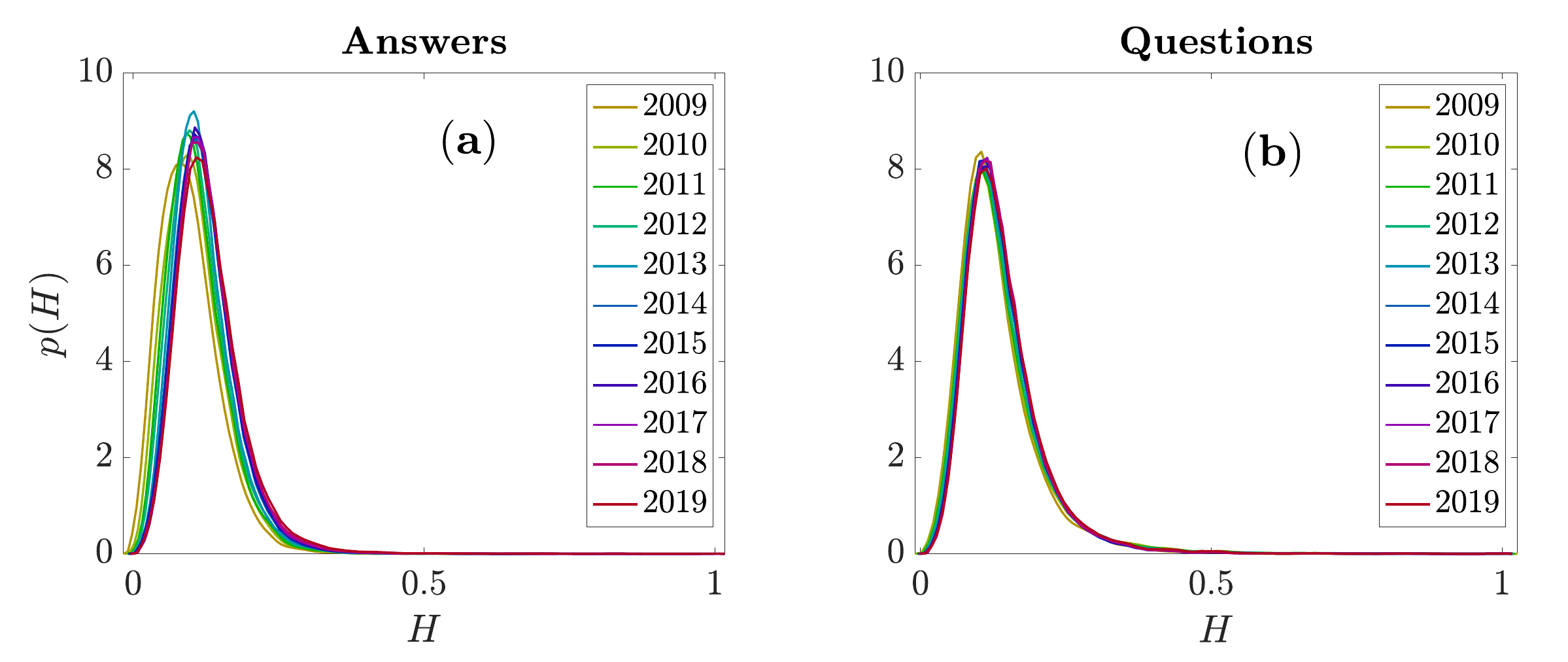}
	\caption{\textbf{Empirical density of the Herfindahl index.} $\mathbf{a}$ Annual distribution of user specialisation with respect to tags in the case of answers. $\mathbf{b}$ Annual distribution of user specialisation with respect to tags in the case of questions.}
	\label{fig:Herfindahl_pdf}
\end{figure*}

\begin{figure*}[h!]
	\centering
	\includegraphics[width=.85\linewidth]{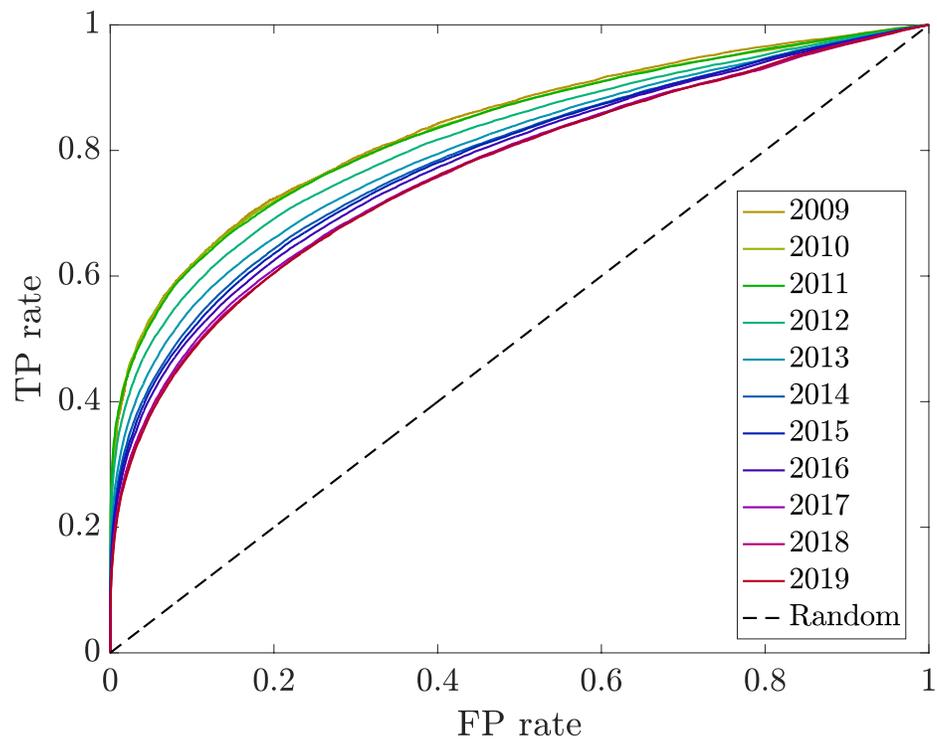}
	\caption{\textbf{Model performance of the logistic regressions for accepted answers}. ROC curves for the models in Table~\ref{tab:regression_acceptance}.}
	\label{fig:ROC}
\end{figure*}

\end{document}